# Agentic End-to-End *De Novo* Protein Design for Tailored Dynamics Using a Language Diffusion Model

Bo Ni[1,2], Markus J. Buehler[1,3,4]*

[1] Laboratory for Atomistic and Molecular Mechanics (LAMM), Massachusetts Institute of Technology, 77 Massachusetts Ave., Cambridge, MA 02139, USA

[2] Department of Materials Science and Engineering, Carnegie Mellon University, 5000 Forbes Avenue, Pittsburgh, PA 15213, USA

[3] Center for Computational Science and Engineering, Schwarzman College of Computing, Massachusetts Institute of Technology, 77 Massachusetts Ave., Cambridge, MA 02139, USA

[4] Lead contact

*Correspondence: mbuehler@MIT.EDU

**Abstract:** Proteins are dynamic molecular machines whose biological functions, spanning enzymatic catalysis, signal transduction, and structural adaptation, are intrinsically linked to their motions. Designing proteins with targeted dynamic properties, however, remains a challenge due to the complex, degenerate relationships between sequence, structure, and molecular motion. Here, we introduce VibeGen, a generative AI framework that enables end-to-end *de novo* protein design conditioned on normal mode vibrations. VibeGen employs an agentic dual-model architecture, comprising a protein designer that generates sequence candidates based on specified vibrational modes and a protein predictor that evaluates their dynamic accuracy. This approach synergizes diversity, accuracy, and novelty during the design process. Via full-atom molecular simulations as direct validation, we demonstrate that the designed proteins accurately reproduce the prescribed normal mode amplitudes across the backbone while adopting various stable, functionally relevant structures. Notably, generated sequences are *de novo*, exhibiting no significant similarity to natural proteins, thereby expanding the accessible protein space beyond evolutionary constraints. Our work integrates protein dynamics into generative protein design, and establishes a direct, bidirectional link between sequence and vibrational behavior, unlocking new pathways for engineering biomolecules with tailored dynamical and functional properties. This framework holds broad implications for the rational design of flexible enzymes, dynamic scaffolds, and biomaterials, paving the way toward dynamics-informed AI-driven protein engineering.

## Introduction

Proteins are not static structure but dynamic molecular machines whose many functions arise from conformational fluctuations across spatiotemporal scales[1]. From an energy point of view, the rugged energy landscape paradigm[2] posits that proteins can sample ensembles of conformations at finite temperature via motions ranging from femtosecond bond vibrations to millisecond domain rearrangements[3]. Such dynamics underpins rich yet essential biological activities and functions, including catalysis, allostery and mechanotransduction. For example, for enzymes such as dihydrofolate reductase and adenylate kinase, transient motions like loop motions and lid fluctuations involving active sites can facilitate the alignment of catalytic residues and the sequestration of substrates[1,4]. Similarly, allosteric mechanisms are often governed by dynamical shits among conformational equilibria triggered by ligand binding thus controlling signal transduction[5,6], such as G-protein-coupled receptor activation through transmembrane helix rearrangements[7]. Among the wide energy or frequency window of dynamics motions, low-frequency vibrations are often crucial in lowering the energy barriers for catalytic reactions[8], facilitating large-scale conformation shift[9], and ligand binding[10,11]. Critically, dysregulation of these dynamics is implicated in disease pathogenesis. For instance, p53 cancer mutants exhibit reduced conformational plasticity, impairing DNA binding[12], while cystic fibrosis transmembrane conductance regulator (CFTR) mutations disrupt gating dynamics essential for ion transport[13]. These observations affirm that the dynamic

"dance" of proteins is not merely a secondary characteristic but a fundamental determinant of their biological function roles. It is essential to understand and engineer proteins with a dynamics point of view.

Over the past decades, a wealth of experimental and computational methodologies has been developed to decode these dynamic phenomena in proteins. Experimentally, techniques such as nuclear magnetic resonance (NMR) spectroscopy[14,15], hydrogen-deuterium exchange (HDX) mass spectrometry[16], cryo-EM[17], single-molecule Förster resonance energy transfer (smFRET)[18], and terahertz spectroscopy[19] have been pivotal in quantifying the time scales and amplitudes of protein motions. For instance, among NMR-based methods[14,20], nuclear spin relaxation rate measurements can report internal motions ranging from subnano- to nano-seconds, while rates of magnetization transfer among protons can capture protein domain movements over milliseconds to days. Complementarily, computational approaches, including molecular dynamics (MD)[21], normal mode analysis (NMA)[22], and elastic network models (ENM)[23], have been employed to investigate the complex motions underlying protein functions. For example, using MD simulations and NMA, it has been demonstrated that the vibrational spectrum and mobility of the spike proteins of coronaviruses can be correlated with the infectiousness and lethality of different variants, thus providing a nano-mechanics approach to estimate the epidemiological effects of new variants[24,25]. While those methods provide some pathways to gain in-depth understanding of dynamics and functions of specific proteins, they are often costly in time and resources, and conventional MD approaches are difficult to scale up. Thus, it remains challenging to connect the functions, dynamics, structures and sequences for a large number of proteins with efficient yet comprehensive ways, and rational engineering and design of proteins based on the desired dynamics properties remain appealing yet rare.

Recent progress in deep learning and generative artificial intelligence (AI) and their applications to proteins is boosting breakthroughs and broadening the horizon of protein research. Tools such as AlphaFold2[26] and RoseTTAFold[27] can predict three-dimensional (3D) atomic structures based on protein sequences with an accuracy comparable to experimental methods but a cost much reduced[28]. Built up on this breakthrough, rapid improvements have been witnessed in reducing computational costs and expanding the applications to orphan sequences[29–32] and protein complexes[33]. However, most of those folding tools are designed primarily to predict static stable conformations and often overlook the intrinsic dynamics that are also pivotal for protein functions. Besides folding predictions, efficient end-to-end models based on deep learning have also been explored to predict protein features in structures (e.g., secondary structures[34–36], binding sites[37] and surfaces[38]), dynamics (e.g., natural vibrational frequencies[39,40]), and properties and functions (e.g., solubility[41–43], melting temperature[44] and strength[45]) for given sequences. Together, these methods and successes provide fast lanes to study the sequence-structure-property relationships in proteins at large scale and encourage the research front expanding towards the more challenging inverse problem of protein design.

Protein design often faces challenges from broad design spaces, limited understanding and formidable costs with conventional methodologies, while deep learning and generative AI bring in new perspectives and possibilities to circulate or overcome many of them[46,47]. For instance, built upon the reliable folding tools and leveraging the creative diffusion models, frameworks like RFdiffusion[48] (All-Atom[49]) and Alphafold3[50] can generate feasible structures and help design *de novo* protein binder, higher-order symmetric architectures and protein complexes with various biomolecules. However, current models often take a rigid picture of the geometry of the designed backbones or functional motifs and lack mechanisms to be conditioned directly by dynamics[51]. It remains rare to design proteins based on the consideration of not only the folded structure but also realistic dynamics[52,53]. At the same time, end-to-end design that skips explicit backbone generation steps have also been explored and shown high efficiency and promise. For example, by merging diffusion model and language models, it has been demonstrated that *de novo* (i.e., not observe in nature yet) protein sequences can be generated based design objectives such as secondary structures[54] or mechanical unfolding responses[55]. Leveraging the general intelligent capabilities of large language models (LLMs) (e.g., GPT-4o[56]), researchers have demonstrated that the workflow of protein design can be automated via LLM-powered multi-agent collaborations[57,58], thus potentially accelerating and scaling up future explorations. It remains unclear whether dynamics-informed protein design can be achieved in an efficient end-to-end manner, which can be particularly suitable for integration with other design goals within a multi-agent multi-modal automated framework.

To address this problem, in this paper, we propose a model composed of protein generation agents that predicts amino acid sequences and 3D protein structures based on key dynamics signature as the design target and aims at achieving accurate yet diverse protein designs. Specifically, in a singular workflow (**Fig. 1**), we start with collecting key dynamics signatures using NMA and full-atom MD for a large number of PDB proteins (**Fig. 1**A). We take the non-trivial low-frequency vibrational mode as the key representation of protein dynamics and focus on the normal mode shape which depicts the heterogeneous distribution of the vibrational amplitude through the backbone. Then, we develop a protein generation model that consists of a protein designer (PD) and a protein predictor (PP) based on protein language diffusion models (pLDMs). The PD is trained to propose amino acid sequences based on the given normal mode shape while the PP learns to predict the normal mode shapes for given protein sequences. At deployment, the two players work collaboratively, mimicking the two-agent team[59], in hope to generate diverse yet accurate designs (**Fig. 1**B). For validation and understanding, we compare the designed sequences with known ones to analyze their novelty, fold and relax the proteins to understand the structural features, and perform NMA using MD models to extract normal modes for design accuracy evaluation (**Fig. 1**C).

Through well-controlled testing and evaluation, we show that our protein generation model can learn complex and degenerated relationships between sequences and normal mode vibrations of proteins in both directions. We demonstrate that at deployment, it can design various sequences based on the given low-frequency normal mode shape, reliably predict their performance on the fly, and result in diverse protein sequences, among which many are *de novo*, that accurately fulfill the desired dynamics design objective. Our two-player framework outperforms the single-model case via collaboratively amplifying the strength of end-to-end models in forward prediction and inverse design and lead to a promising synergy of diversity, accuracy and novelty for protein design tasks. Combining these results and the essential roles of dynamics in protein functions and performance, we believe our end-to-end dynamics-informed protein design agent model and similar frameworks can provide novel navigating tools for gaining in-depth understating of sequence-structure-dynamics-function relationships of proteins in the collective level and open fast lanes for challenging protein design tasks that require various dynamics signatures as conditions. We expect our models will be useful in numerous biological and engineering applications for the dynamics-sensitive function-targeted generative design of various proteins and protein materials.

## 2. Results and discussions

### Protein database on low-frequency vibrational normal modes

While dynamics of proteins can involve different spatiotemporal scales, low-frequence modes present key yet efficient signature for dynamics that are essential to many biological processes and functions of proteins. From a mechanics point of view, these low-frequency modes are the motion patterns with low deformation energy penalty and mainly contribute to the flexibility of proteins[60,61]. It has been demonstrated that by analyzing low frequency modes, researchers have gained insights into of protein motions involved in ligand binding[62], confirmational changes[63], enzyme catalysis[64] and protein-protein interactions[65].

Therefore, here we adopt low-frequency normal modes to represent the essential signature of protein dynamics. To cover the detailed relationship between sequences, structure and dynamics, we focus on the vibrational displacement distribution in protein molecules instead of the frequencies[39,66,67]. Following the previous study[66], we use full-atom MD with the CHARMM[68] force field to relax the protein structure. Then, NMA for protein structure can be performed by solving the eigen value problem of the Hessian matrix, whose components are the second-order partial derivatives of the potential energy function of the fully atomic model with the adopted force field[68]. While the first six modes are trivial with zero frequency for rigid body motions, we focus on the nontrivial ones starting from the seventh. As a model study, here we only sample the first non-trivial normal mode with the lowest frequency for the following study. It should be noted that, with our method and models, it is straightforward to expand and cover other non-trivial low frequency modes. More details on the NMA calculations using MD models of proteins can be found in the **Materials and Methods** section.

As shown in **Fig. 2**A, the displacement field of the lowest non-trivial normal mode of an example protein monomer is depicted by the solid vectors in red. To sample this vibrational displacement field, we collect the displacement components (dash lines in **Fig. 2**B) of $C_\alpha$ atoms of the residues through the backbone from the N-terminal to the C-terminal. We observe that the distribution of this displacement field is heterogeneous along the backbone and can be related with the local structure and flexibility. For example, the two ends (terminals N and C in **Fig. 2**A and B) with relatively loose and open geometry show relatively larger vibrational displacement compared to residues within the compact geometry inside the backbone. Away from the terminals, the protein consists of segments of alpha helix connected with hydrogen bonded turns and coils. The residues with turn or coil-type secondary structure (marked as P in **Fig. 2**A) are expected to be less confined and more flexible than those in alpha helices. Correspondingly, a local maximal of vibrational amplitude is observed around this portion inside the backbone. To represent vibration details along the backbone as exemplified above, we use the amplitude of the vibrational displacement to define a normal mode shape vector, $\vec{V}_A$, for a protein monomer with $N$ amino acids as the following,

$$\vec{V}_A = [d_1, d_2, \dots d_i, \dots, d_N] \tag{1}$$

where $d_i = \sqrt{d_{ix}^2 + d_{iy}^2 + d_{iz}^2}$ for $i = 1,2, \dots, N$, and $[d_{ix}, d_{iy}, d_{iz}]$ are the 3D displacement vector of the normal mode sampled at the $i$-th residue. Since normal mode vector can be scaled arbitrarily, to compare proteins of different sequence lengths, we normalize the normal mode shape vector such that,

$$\|\vec{V}_A\| = \sqrt{\sum_{i=1}^{N} d_i^2} = N \tag{2}$$

where $\|\cdot\|$ is the operator to calculate $L_2$ norm of a vector and $N$ is the sequence length of the protein. It should be noted that unlike the displacement components, this normal mode vector $\vec{V}_A$ of displacement amplitude is independent of choice of coordinate systems, thus being an invariant descriptor of the normal mode vibration.

To curate a dataset for naturally existing proteins on their dynamics signatures using low-frequency normal modes, we apply the protocol above to proteins with a sequence length no more than 126 amino acids from the Protein Data Bank (PDB)[69] dated by Jan 11, 2024 using an automated parallelize workflow similar to the previous work[66]. Results of 12,924 protein monomer chains are collected. Further details of the dataset can be found in the **Materials and Methods** section. An overview of the distributions of the normal mode information are shown in **Fig. 2**C-D and **Fig. S1**. Specifically, in **Fig. 2**C, the normalized mode shape vectors show peaks at various locations, indicating the complexity embedded even in the lowest non-trivial vibrational mode across different proteins. **Fig. 2**D shows the distribution of the residues that undergo the maximal vibration amplitude in the backones. Two peaks at the open ends suggest a common trend that two terminals tend to have strong vibrations. The distributions of sequence length and normal mode frequences can be found in **Fig. S1**. An in-depth analysis of the normal mode distribution and its relationship to protein structures, flexibility and sequences may reveal important insights on the statistical scale and deserve a separate study in the future work. Further insight can also be obtained by combing the dataset with numerous experimental studies using NMR and other techniques. Here, we focus on applying this newly collected data to develop generative models. Next, we develop generative AI models, in hope to link the protein sequences and normal mode shape vectors bidirectionally and generate proteins based on the given normal mode shapes, and evaluate the accuracy, diversity and novelty of the designs.

**Agentic protein generation model and inverse design for normal mode shapes**

Previous works have demonstrated that the protein language diffusion models (pLMDs)[55] can combine the deep knowledge of protein sequences baked in the pretrained protein language models[30] and the learning and designing capabilities of diffusion models[54,70] to map complex conditions (e.g., nonlinear mechanical unfolding responses) to protein sequence space. At the same time, the generating tasks based on the lowest non-trivial normal mode shape studied here present unique challenges in complexity and degeneracy. On one hand, as exemplified in the individual cases (e.g., **Fig. 2**A and B), the normal mode shape is determined by detailed 3D geometry of backbone, hierarchy structures as well as elasticity of protein and can be sensitive to both local (e.g., secondary structure type) and global features (e.g., topology of the backbone), which makes it an non-trivial task to directly link sequences with normal mode vibrations. On the other hand, based on insight from mechanics, information of the single normal mode is clearly not sufficient to specify the whole system (e.g., the Hessian matrix of the protein). And the chosen normal mode vector consisting of displacement amplitudes further loses the directional information of the original vibration. Thus, the probability of finding proteins of different structures as well as sequences but sharing the same or similar normal mode shape vectors can be high, which leaves the inverse design problems highly degenerative and introduces interesting possibilities to understand proteins from a perspective of classes of designs that relate to certain set of viable dynamical behaviors.

To address these challenges, here we invoke two separate pLMD models, a protein designer (PD) and a protein predictor (PP), to learn the forward prediction and inverse design tasks between sequence and normal mode spaces and organize them as collaborative agents to address the protein generation tasks based on dynamics signatures of normal mode shape. As shown in **Fig. 3**A, the PD is tasked with generating sequences based on the given dynamic property. It combines a protein language model (pLM) pretrained on large sequence corpora (shaded in orange in the right) and a trainable diffusion model built with one-dimensional U-Net architecture with attention mechanisms (shaded in pink at the middle). The pLM is tasked with translating proteins between the token space and its pretrained latent space. While the diffusion model learns to sample and improve new points in such space based on the conditioning encoded from dynamic property (shaded in purple) via multiple challenges (E1 and E2) during the denoising process. The PP in **Fig. 3**B processes similar components but aims to predict the dynamic property for the given sequences. During the denoising process, the prediction is gradually improved under the conditioning using multiple representations of the given protein sequence via the frozen pLM, including the hidden state (R1) and the softmax probability based on the logits (R2). The result is then translated back into the dynamic property space via a decoder (D). We train the two models separately. More details about the model and training can be found in the **Materials and Methods** section.

To boost performance at deployment, we borrow inspiration from the multi-agent frameworks and organize the PD and PP as a collaborative agentic system. As shown in **Fig. 1**B, for a given design objective of normal model vector, the PD is tasked to generate ensemble of sequences as candidates. On the spot, the PP will predict their normal model vectors, thus evaluating the performances of the generated batch. Depending on the demand, results that prioritize accuracy or diversity can be screened. For cases where the demand is not satisfied, iteration of the previous steps can be invoked.

We test the performance of our protein generation model using the normal mode shape vectors of the proteins from the test set, with which the models have not been trained. Here, we start by looking for the most accurate design. To do so, for each design goal, the PD designs 40 candidates, from which the PP selects the best one based on the accuracy it predicted. We then validate the generated sequences using the same NMA protocol introduced in the previous section. Besides, the folded 3D atomic structures of the generated sequences are predicted using OmegaFold[71]. With protein BLAST[72] test and DSSP[73], we exam the novelty of the generated sequences, identify their secondary structures and discuss potential relationship with normal mode shape.

**Fig. 4** shows some examples of the designed proteins and their normal mode shapes measured using our protocol. In terms of the design objective, the input normal mode shapes as condition (red curves) in **Fig. 4**A-F covers a variety of representative patterns, including a L-shape case (A) with the maximal vibration concentrated near the N-terminal and relatively weak amplitude at other positions along the sequence, a horizontally flipped L-shape (B) with the maximal vibration occurring at the C-terminal, a U-shape (C) with both open ends, N- and C-terminals with strong vibration amplitude surpassing other portion of the backbone, a W-shape with strong vibrations at both terminals and the middle region separated by two nearly zero stationary nodes in between, and cases with single (E) or multiple (F) internal peaks of strong vibration surpassing the open ends. Note that on top of these simplified shapes, these realistic design objectives also include relatively small but complex oscillations. Despite such variety and complexity of normal mode shapes, the proteins generated by our model demonstrate measured normal mode shapes (blue curves) that in general closely follow the design objectives. We use multiple metrics, including the Pearson coefficient, $\rho$, and relative $L_2$ error, $L_2^{rela}$, (see **Materials and Methods** section for details), to measure quantitatively the accuracy of the design in filling the design objective of normal mode shape. The relatively large $\rho$ (between -1 and 1) and small $L_2^{rela}$ listed in **Fig. 4** indicate our generation agents can produce accurate design for these various design objectives of normal mode vibration.

Corresponding to the various patterns of the vibrational amplitude of the lowest non-trivial mode, the generated proteins also show a variety of geometry and internal structures, some of which may be related to the vibrational motion. For instance, in cases B-C, the unstructured coils are often observed at the region near the open ends with concentrated strong vibration amplitude (C-terminal in case B, both terminals in case C). In comparison, more compact backbone geometry and secondary structures with stronger confinements (e.g., H-bonds in alpha-helix and beta-sheet) can suppress vibrations (like middle regions in cases A-C). Even for regions near the open end of backbone, by adopting confined secondary structures (e.g., alpha-helix), the relative vibration amplitude can still be suppressed (e.g., N-terminal in case B). Similarly, comparing the beta-sheets with organized overlapping and the relatively less confined connections between beta-sheets, the latter often contribute to higher vibration amplitude in the middle of the sequences (e.g., P in case E and $P_1$-$P_4$ in case F). In case D, the relatively short sequence takes a continuous alpha-helix structure, which can be approximated as an elastic rod stabilized by hydrogen bonds. The observed vibration shape agrees with this approximation in terms of the lowest normal mode shape.

We also compare the generated protein sequences with the known one from the test set used to provide the design objectives and find relatively low recovery ratio (see the **Materials and Methods** section for details) of animo acids along the sequences, which indicates the generated sequence can be different from the known ones in the test set. To investigate the novelty of these generated proteins, we apply basic local alignment search tool (BLAST)[72] analysis to the predicted amino acid sequences to access whether, and to what extent, they represent *de novo* sequences or closely related forms of known proteins. **Table 1** shows the results of the BLAST analysis for the various cases listed in **Fig. 4**. We find that even though the input design targets are from existing PDB proteins, many of the generated protein sequences (cases shown in **Fig. 4**B-E) do not match any sequences in the database of known proteins with standard BLSAT analysis[74] (i.e., returning "no significant similarity found" in protein BLAST test) and are *de novo* ones. The model can also produce sequences (e.g., cases A and F in **Fig. 4**) that show some similarity to the existing proteins. While the model is only trained on a small portion of PDB proteins, with the normal mode shapes of existing PDB proteins as an input and considering the possible degeneracy, we expect the possibility of the model "rediscovering" sequences that show some similarities to the known proteins. Further measures may be utilized to boost the novelty of design for such cases, including screening sequences based on BLAST results. It should also be noted that, given such novelty in the generated sequences, the normal mode shapes predicted on the spot by the PP (green dash curves in **Fig. 4**) still reasonably agree with the measured ones (blue curves in **Fig. 4**), indicating that the PP remains reliable for *de novo* sequences generated by our PD.

Besides focusing on individual cases, we also show the distributions of the design accuracy and novelty for a larger number of testing cases. **Fig. 5** summarizes the results of 1,293 proteins generated based on various normal shape vectors from the whole standalone test set. On the normal mode shape, the Pearson coefficient $\rho$ between the measured normal mode shape vectors and the input conditions among cases (in blue in **Fig. 5**A) show unimodal distributions with the highest peak of population near 0.87, indicating cases with satisfying accuracy. However, the distribution also covers a broad range (between 1.00 and -0.50) with a median of 0.53 and decays slowly towards the region of poor accuracy, indicating there also exist cases of relatively poor accuracy. The distribution of the relative L2 error with a median value of 0.57 (blue data in **Fig. 5**B) also indicates a limited accuracy. These observations reflect the intrinsic difficulty in solving such protein design problems with high residue-level accuracy. Indeed, shown in **Fig. 5**C, the component-wise comparison of all normal mode shape vectors concentrates around y=x line with a broad span and the Pearson coefficient reached (0.51) is close to the median for the vector-wise value (0.53 in **Fig. 5**A). However, as discussed above, for vibration-based design of proteins, the overall shape of the normal mode may attract more interest than small oscillations localized to residues. For example, the overall shapes (e.g., L, U, W shapes) of normal mode vectors discussed in **Fig. 4** may prove to be more relevant to applications such as protein binder design than the small oscillations on top of them. Thus, there exists rationale to filter the original normal mode shapes and investigate the accuracy in terms of the smoothed version. To do so, we apply a low-pass filter to the measured and conditioned normal mode shape vectors using fast Fourier transformation (FFT) and then compare them. The low-pass filter adopted allows the contributions from the lowest 10% frequencies to pass while removing others. The smoothed normal mode shape vectors often maintain the overall trend of the original data while free of small oscillations (see examples in **Fig. S2**). With such smoothed normal mode shape vectors, the corresponding Pearson coefficient and relative L2 error (in red in **Fig. 5**A and B) distributions shift clearly towards the high accuracy region and achieve improved median values of 0.72 and 0.37 respectively. This shift demonstrates that our agentic protein generation model can achieve higher accuracy on the overall shape of the normal mode vectors for a large number of cases. Moreover, this contrast suggests that our method more reliably captures the large-scale (low-frequency) portion of the mode shape but is less precise on the finer, residue-by-residue details. Our framework appears to be particularly successful at reproducing overall vibration "profiles," which are the most biologically relevant for large-scale conformational dynamics.

On the novelty of the designed proteins, **Fig. 5**D shows a bimodal distribution of the highest percent identity found via protein BLAST analysis for all the generated sequences. The highest peak (on the left in **Fig. 5**D) corresponds to the cases where the generated proteins have little similarity to the existing/known ones and are totally *de novo*. There also exists the other weaker peak on the right for cases in which the proteins generated are similar to known proteins. The bimodal distribution echoes the result of individual cases listed in **Table 1** and the relative height of the two peaks indicates our model has a stronger tendency in generating *de novo* sequence designs. We conclude that our approach effectively explores protein sequence space well beyond evolution's "comfort zone," significantly expanding the repertoire of possible structural and dynamic solutions.

**Benefits of using protein generation agents in boosting design diversity and accuracy**

To investigate the potential of our protein generation model in capturing possible design diversity, we sample the top 4 sequences from the 40 candidates designed by the PD based on the prediction of the PP. **Fig. 6** shows examples for a U-shape normal mode vector and a L-shape one. For the U-shape input, the 4 sequences, U1-U4, designed by our agentic model all achieved high design accuracy as the measured normal mode shapes follow closely with the condition (**Fig. 6**A). However, the 3D structures of the 4 proteins show clear differences as well as similarities, which can be related to the prescribed normal mode shape. As shown in **Fig. 6**B, the designed proteins all adopted a relatively compact core region with two open ends expanding out, corresponding to large vibrations near the ends and limited vibrations in the middle. The regions near the two terminals share similar secondary structures of unstructured coils (highlighted in green and red for N- and C- terminal respectively in **Fig. 6**C) and extend away from the compact core. While the compact core parts show a variety of secondary structure types, including bundles of alpha-helices (U1 and U2) and mix of alpha-helix and beta-sheets (U3 and U4 in **Fig. 6**C). Correspondingly, the amino acid sequences (U1-U4 in **Table 2**) also show some similarity near the two ends and keep diversity for the middle parts. A similar pattern can also be observed for design cases (L1-L4) with an L-shape condition as shown in **Fig. 6**D-F and **Table 2**.

Combining these observations, it suggests that for backbone regions to achieve relatively high and concentrated vibration amplitude, it often requires less confined coils or turns with limited options of secondary structures. While multiple choices of confined structures exist for regions prescribed with suppressed vibration amplitude, ranging from alpha-helix, beta-sheet to their various mix. Some diversity of designs based normal mode shapes can come from such various choices in structures and sequences for the suppressed region, and our model appears to capture such degeneracy and come up with a range of designs. As shown in Table 2, surprisingly many of the multiple designs based on the same input conditions still do not find similarities among the known proteins and are *de novo* (U1, U2 and U4 for the U-shape design and L1 and L2 for the L-shape design). Combining these results, it has demonstrated that our design approach can achieve the synergy of accuracy, diversity and novelty for dynamics-informed protein design with suitable design conditions. It should also be noted that the achievable diversity of our model can be affected by the choice of the input normal mode shape. As shown in **Fig. S3**, with the multi-peak shape (**Fig. 4**F) as the input normal mode shape, the designed proteins present very similar secondary structures (i.e., multiple beta-sheet connected by turns and coils shown in **Fig. S**B-D) Thus, the diversity in potential protein sequences may also be limited, especially in the regions near the peaks (**Fig. S3**E).

To investigate the effect of the PP in improving the design accuracy, we sample both the best and the worst designs according to the PP from the 40 candidates proposed by the PD. The comparison of design accuracy in terms of Pearson coefficient of those groups, the predicted best (in blue) and the predicted worst (in red), on the whole test set is shown in **Fig. 7**A. The former shows a distribution with the main peak at the high accuracy region (near 1) while the latter group peaks in a low accuracy region (near 0). Similar relative rank can also be observed in terms of their median values (0.53 vs 0.31). Thus, the predicted best group does outperform the predicted worst, and the PP distinguishes them correctly on the collective scale. At the same time, the predicting accuracy of the PP on the two groups shows little difference (**Fig. 7**B), indicating the PP maintains reliable performance on protein sequences with different design accuracy. Given the clear gap between the worst and the best groups which are all designed by the same PD, it becomes clear that the integration of the PP during the deign process is essential to improve the design accuracy while avoiding the high cost of invoking physics-based tests.

Looking at the results of our experiments, we further note that the model appears to leverage secondary structure elements to tune local flexibility, confirming that it "understands" the relationship between backbone hydrogen-bonding motifs and vibrational amplitude. We can see this, for instance, in **Fig. 4A-F**, where regions predicted to have low amplitude, we often see more confined secondary structures (e.g., α-helices or β-sheets), whereas in higher-amplitude regions, such as loop segments or chain termini, the structures are more open or coil-like). In **Fig. 6B-C and E-F** we present a side-by-side comparison of four designs generated for the same target mode shape. These panels color-code the predicted secondary structures for each design, illustrating a clear tendency for α-helices or β-sheets to populate lower-amplitude backbone regions, while loops and coils emerge in higher-amplitude segments. This pattern highlights how the model captures the relationship between secondary structure motifs and local flexibility, using specific structural elements to tune vibrational amplitude along the protein chain.

## 3. Conclusion

In summary, we have introduced a novel, dual-component protein language diffusion framework, consisting of a forward and inverse model, which bridges sequence generation with vibrational dynamics prediction to achieve *de novo* protein design. By conditioning sequence generation using the generative inverse design model on specified normal modes of vibration and rigorously screening candidates for dynamic fidelity, our approach substantially boosts design accuracy, diversity, and novelty, thus transcending the limitations of traditional static design paradigms. Incorporating the PP as a second agent in our agentic two-step workflow raises the average correlation coefficient by filtering out designs that deviate from the target shape. This synergy is a major reason our final designs show robust performance, as the PP effectively corrects for the inherent stochasticity of the PD model while reflecting an agentic approach that iterates between generation and validation.

Our results demonstrate that proteins designed via this generative agentic model not only fold into stable, novel structures but also reproduce targeted vibrational amplitude profiles along their backbones. This establishes a direct, end-to-end linkage between sequence and dynamic behavior, offering a powerful route to engineer proteins with bespoke functional dynamics. In doing so, our work complements recent breakthroughs in static structure prediction[26] and generative design[48], pushing the envelope toward a more complete understanding of protein functionality that includes the essential role of dynamics. When we target a single normal mode shape, we often observe multiple top candidates that differ significantly in primary sequence yet converge on similarly accurate normal mode profiles. This underscores that designing for a single vibration shape does not necessarily fix the backbone or sequence, and that our work provides evidence of a large degeneracy in sequence space. This observation suggests that low-frequency normal modes alone do not pin down a unique sequence or fold but instead correspond to a family of viable solutions. it appears alpha helices are often used to suppress local vibrational amplitude, whereas beta sheets plus interspersed coils can produce more internal peaks. Looking a bit deeper into the results, we note that the model appears to leverage secondary structure elements to tune local flexibility, suggesting that it "understands" fundamental relationships between backbone H-bonding patterns and vibrational amplitudes, as can be seen in **Figs. 4** and **6**. This provides evidence for an important link between structural motifs (α-helix, β-sheet, coil) and dynamic patterns (low vs. high amplitude).

Looking forward, several avenues merit further exploration. First, it is straightforward to expand our mode to include more dynamics information as input condition, including normal mode frequencies[66], directional information of normal mode shape, and multiple non-trivial modes. It remains open and interesting to investigate how such detailed conditions will affect model performance and the diversity in the design. Second, integrating our AI-driven models with other end-to-end models[39] as well as physics-based approaches through LLM powered multi-agent automated frameworks[57,75] may enhance the predictive power, mechanistic interpretability of dynamic behaviors and efficiency in design. Third, comprehensive experimental validation, using techniques such as NMR spectroscopy, terahertz spectroscopy, or single-molecule FRET, will be crucial to confirm the in-silico predictions and assess the functional impact of engineered dynamics in cellular contexts[20]. Finally, while our framework has successfully expanded the accessible protein sequence space, challenges remain in capturing the full complexity of multi-scale dynamic phenomena and translating these insights into predictable biological outcomes.

By uniting advanced generative AI methodologies with deep biophysical insights, our work lays a robust foundation for the rational design of proteins that harness dynamic vibrational properties as a functional design parameter. This integrative approach opens new horizons for the development of enzymes, sensors, and biomaterials with unprecedented capabilities, charting a path forward in the evolving landscape of protein engineering.

## 4. Materials and Methods

### Normal mode analysis of PDB protein models in molecular dynamics

Following the previous work[66], we downloaded the protein structures from the Protein Data Bank. The atomic structures are cleaned, separated and completed to get the individual polypeptide chains using Visual Molecular Dynamics (VMD)[76], Multiscale Modeling Tool (MMTSB) toolset[77], and SCWRL4[78]. Then, the protein chain structures are relaxed via energy minimization based on the CHARMM19 all-atom energy function and an implicit Gaussian model for water solvent[79,80]. Before the NMA, 10,000 steps of energy minimization with a steepest descent algorithm and another 10,000 steps of energy minimization with an adopted basis Newton-Raphson algorithm are performed for further relaxation. We use the Block Normal Mode (BNM) method[81,82] in CHARMM for NMA of each protein chain for high efficiency. We save the results of eigen values and eigen vectors of the normal modes of interest. More details can be found the previous work[66].

### Dataset

We curate the dataset based the NMA results. Key information for each protein case includes PDB ID, protein sequence, sequence length, normal mode frequency, normal mode shape vector, the index of the residue with the maximal vibrational displacement of the normal model. See **Fig. 2** and **Fig. S1** for their distributions and the dataset in **SI** for complete data. For training, we randomly pick 90% of the dataset as the training set and set the remaining 10% aside for testing.

### Design of the architectures of deep learning models and training

Both the PD and PP are protein language diffusion models (pLDMs)[55] consisting of a pretrained protein language model (pLM) and a diffusion model. Only the latter is trainable. There are multiple choices for the pretrained pLM and usually larger pLMs require higher computing resource and cost. To balance computational efficiency and performance, we adopt a medium-sized pretrained model with 150M parameters from the ESM-2 series based on the previous study[55]. In the diffusion model, the condition is integrated into the denoising process via multiple challenges of the U-net, including as the partial input for the denoising time step and concatenation with middle results. We train the two models separately using an Adam optimizer and setups similar to the previous works[54,55].

**Protein folding**

We adopt OmegaFold[71] for rapid prediction of protein structures from the sequence. OmegaFold offers a rapid alternative as it does not require Multiple Sequence Alignment (MSA) yet produces results of similar accuracy as AlphaFold2[26] and trRosetta[83] (and similar, related state of the art methods).

**Design accuracy evaluation**

We use various metrics to compare the measured normal mode shape vectors with the input design conditions for individual designs as well as predictions for the whole test set.

For vectors, including the normal mode shape vector for one protein and its components of all proteins in the test set, the Pearson coefficient $\rho$ and relative L$_2$ error $L_2^{rela}$ defined as the following,

$$\rho[\vec{x}, \vec{y}] = \frac{\sum_i (x_i - \bar{x})(y_i - \bar{y})}{\sqrt{\sum_i (x_i - \bar{x})^2 \sum_i (y_i - \bar{y})^2}} \tag{3}$$

$$L_2^{rela}[\vec{x}, \vec{y}] = \frac{\|\vec{x} - \vec{y}\|}{\|\vec{x}\|} = \frac{\sqrt{\sum_i (x_i - y_i)^2}}{\sqrt{\sum_i (x_i)^2}} \tag{4}$$

where $\vec{x}$ is the ground truth or input vector and $\vec{y}$ is the measured one from the predictions, $x_i$ and $y_i$ are their components and $\bar{x}$ and $\bar{y}$ are the means of the components $x_i$ and $y_i$.

To compare the generated protein sequences with the one used to provide the input normal mode shape vector, we define the recovery ratio of the generation as the following,

$$\text{Recovery ratio} = \frac{n}{N} \tag{5}$$

where n is the number of the residue in the generated sequences with the same amino acid type with the known sequence sequences from the test set and $N$ is the sequence length. This recovery ratio is between 0 and 1.

**BLAST analysis**

The basic local alignment search tool (BLAST)[72] analysis for the various cases is conducted using the blastp (protein-protein BLAST) algorithm[74], and the non-redundant protein sequences (nr) database.

**Visualization**

We use Visual Molecular Dynamics (VMD)[76] for visualization of the protein structures.

**Software versions and hardware**

We use Python 3.10.13, PyTorch 2.3.1+cu13[84] with CUDA (CUDA version 12.4), and a NVIDIA Tesla V100 with 32 GB VRAM for training and inference.

**Acknowledgments:** We acknowledge support from USDA (2021-69012-35978), the MIT-IBM Watson AI Lab and MIT's Generative AI Initiative.

**Conflict of interest**

The author declares no conflict of interest.

**Data and materials availability**

All data needed to evaluate the conclusions in the paper are present in the paper and/or the **Supplementary Materials**.

Codes and model weights are available at https://github.com/lamm-mit/ModeShapeDiffusionDesign and https://huggingface.co/lamm-mit/VibeGen.

**Author contributions:** MJB and BN conceived the study. BN curated the dataset, developed and trained the neural network and performed associated data analysis and prepared the first draft. MJB supported the analysis and wrote the paper with BN.

## Supplementary materials

Additional figures, PDB files, and other materials are provided as **Supplementary Materials**.

- A CSV file of the **curated dataset** on protein sequences and normal modes used for training and validation cases.
- A ZIP file with **protein structure** PDB files for the proteins generated by the model with some representative normal mode shape vectors (Fig. 4 and 6).
- Movies for the lowest non-trivial normal mode vibrations of selected protein designs (Fig. 4 and 6).

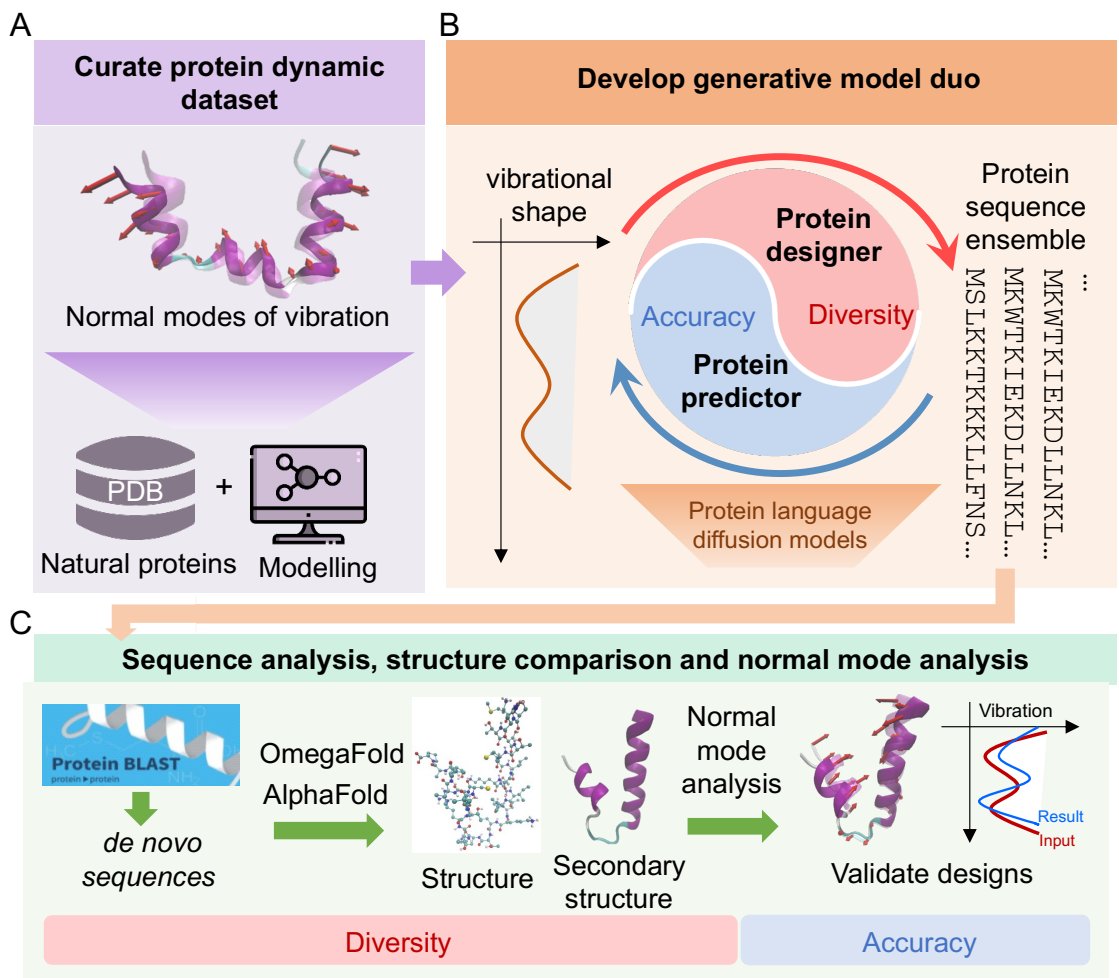

**Fig. 1. Workflow of developing the end-to-end protein generation model based on dynamics signature, featuring an agentic system of two models that collaborate to solve the problem.** (A) curating a PDB protein dataset on their nontrivial low-frequency vibrational normal modes as their dynamics signature. (B) Overview of the protein generation mmodel based on protein language diffusion models. The agentic model consists of a protein designer (PD) and a protein predictor (PP). The PD proposes various protein sequences based on the given vibrational shape of the vibrational normal mode and boosts diversity in the design. While the PP predict the normal mode shapes for the given protein sequences to evaluate the accuracy. During the generation deployment, the two components work together mimicking a two-agent team to design and screen sequences, thus achieving the balance of accuracy and diversity for the generated sequences. (C) Analyzing and validating the generated proteins. The protein-protein BLAST test is employed to analyze the generated sequences and screen for the de novo ones. Folding tools like OmegaFold and AlphaFold2 are used to predict the atomic structures of the sequences. And the secondary structures are analyzed. Using molecular dynamics and normal mode analysis, the vibrational shape of the low-frequency normal modes of the generated proteins are obtained and compared with the input design objectives to validate the design.

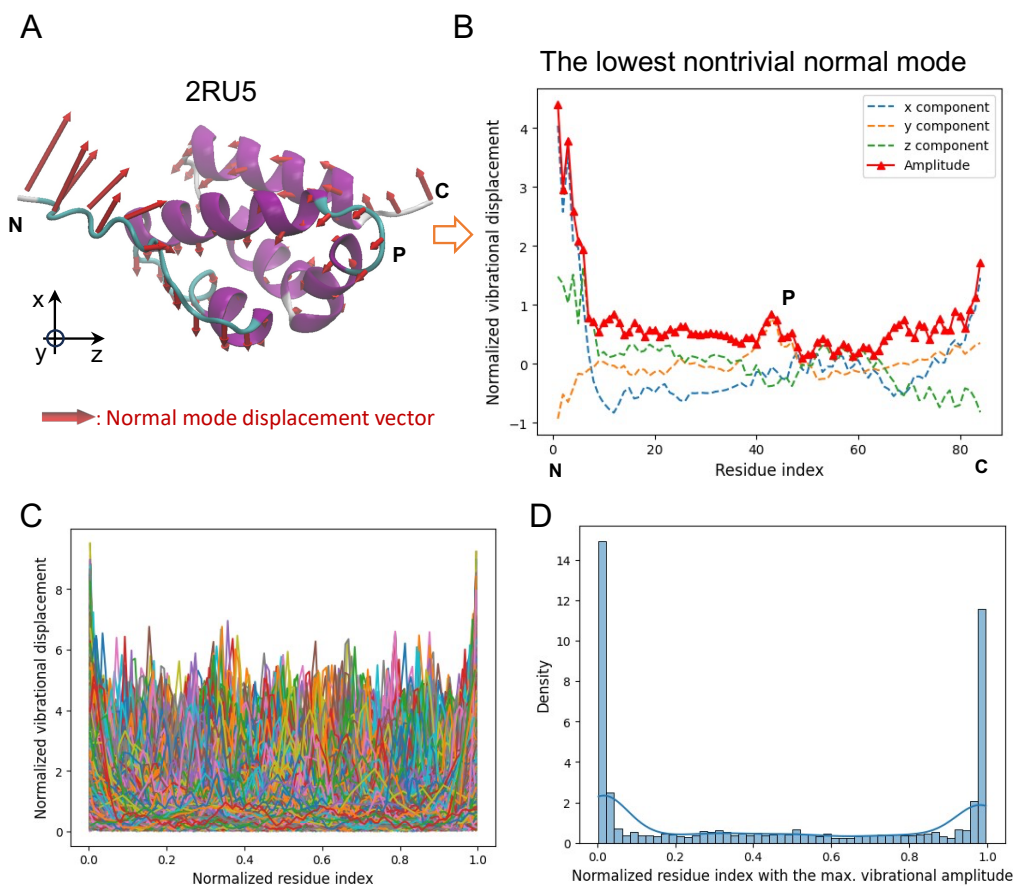

**Fig. 2 Normal mode analysis of proteins and low-frequency mode shape dataset curation.** (A-B) The lowest non-trivial normal mode of a PDB protein obtained using normal mode analysis and full-atom molecular dynamics model. Red arrows in (A) represent the displacement vector of this normal mode. In (B), the components and the amplitude of the vibrational displacement across the backbone are collected at the $C_\alpha$ in each residue. The distributions of the displacement are heterogeneous along the backbone (B) and sensitive to the local structure and flexibility of the protein (A). The vector of this lowest non-trivial normal model displacement amplitude is termed as the normal mode shape vector to represent the dynamics signature of the protein. (C) Collecting the normalized normal mode vector for a large number of PDB proteins. (D) the distribution of the residue with the largest vibrational displacement amplitude. In (D) and (C), the indices of residues are normalized between 0 and 1 for the convenience of comparison among different proteins.

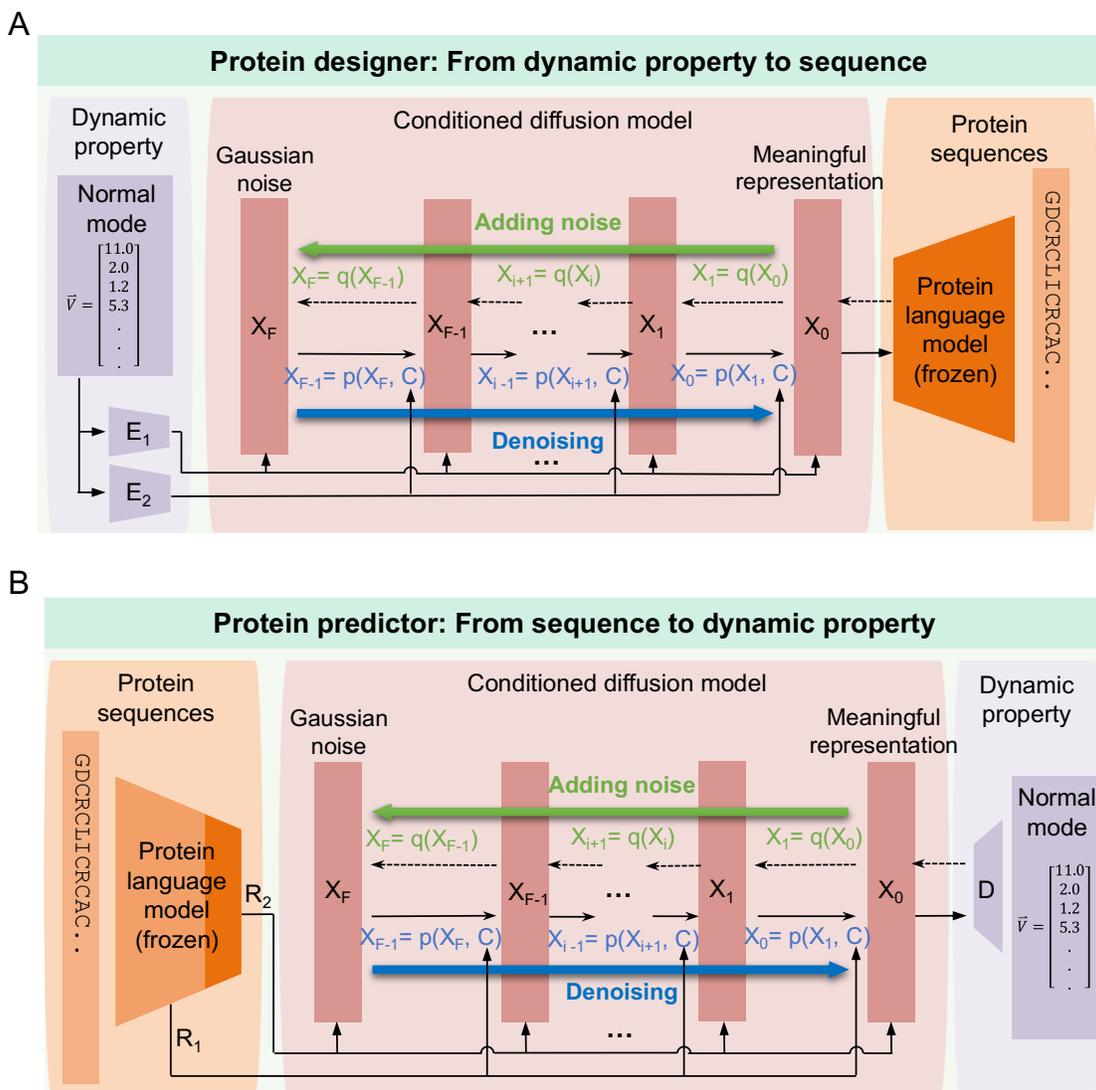

**Fig. 3. Overview of the structures of the protein generation model.** (A) Structure of the constructed protein designer that generates protein sequences based on the given dynamic property. It combines a protein language model pretrained on large protein sequence corpora (shaded in orange) and a trainable diffusion model (shaded in pink). During the denoising process, the generated sequences are conditioned via multiple channels (E1 and E2) mapped from the dynamic properties (shaded in purple). E1 and E2 are trainable encoders. (B) Structure of the designed protein predictor that predicts the dynamic property of the given protein sequence. During the denoising process, the prediction is conditioned by different representations of the sequences via the pretrained protein language model, including the hidden state (in R1 channel) and the softmax probability based on the logits (through the R2 channel). D is a trainable decoder for normal mode shape vectors.

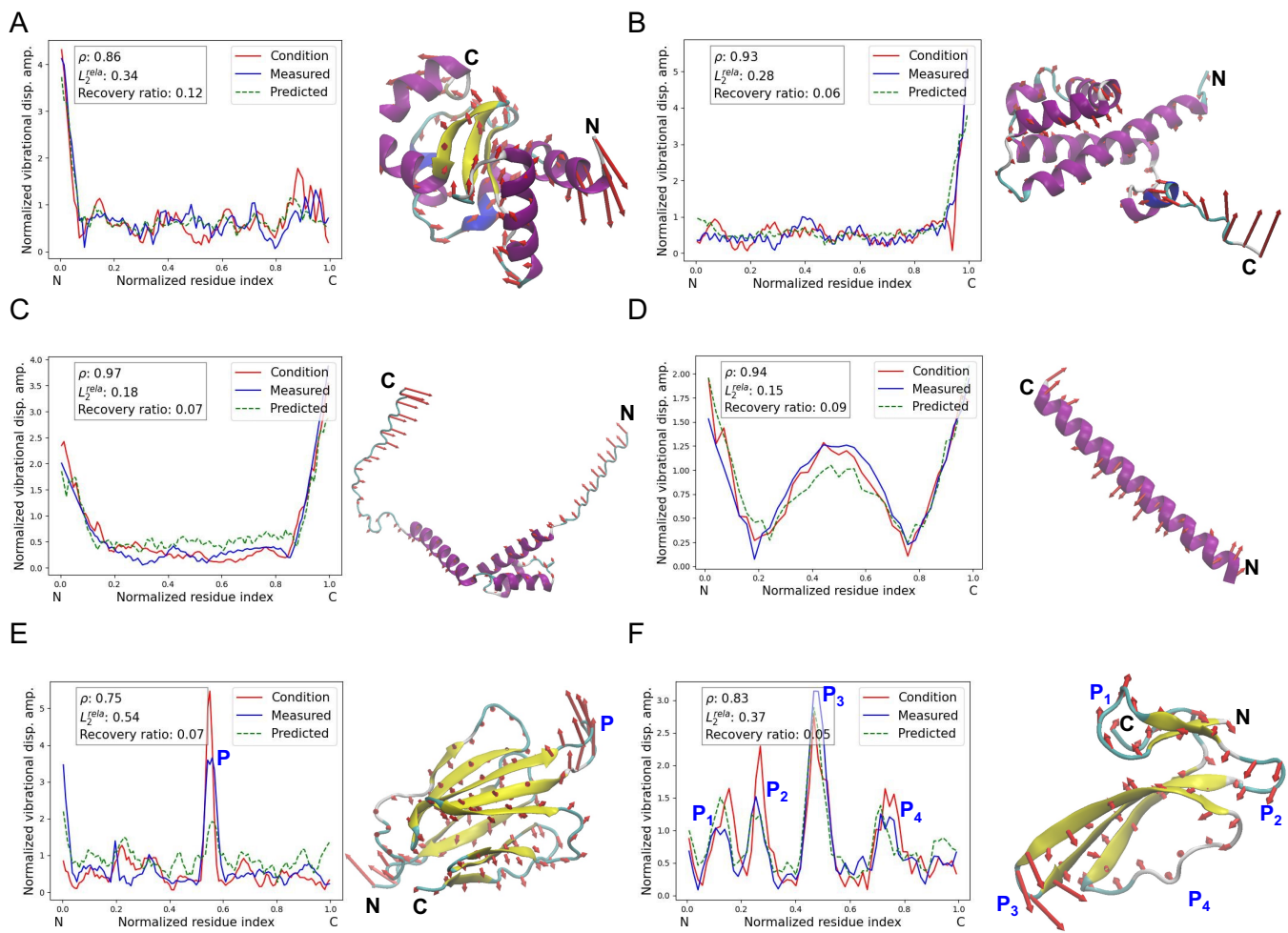

**Fig. 4. Results for protein generation based on the normal mode shape vectors of naturally existing proteins.** Panels A-F show a variety of representative cases of different normal mode shapes (red curves), including an L-shape case (A) with high vibration amplitude near the N-terminal, an flipped L-shape (B) with high vibration amplitude near the C-terminal, a U-shape (C) with large vibration near both terminals, N and C, and a W-shape (D) with two stationary nodes and strong vibrations at both the open ends and in the middle of the sequence, a case (E) with a single localized peak (P) of vibration away from the terminals, and a case (F) with multiple internal peaks (P$_1$-P$_4$) of large vibration. The proteins generated by our model have demonstrated normal mode shapes (blue curves) that follow the trend of design objectives (red curves). Given the complexity and oscillating nature of the normal mode shapes, we use Pearson coefficient $\rho$ and relative L2 error $L_2^{rela}$ to measure the accuracy of the design. We also compare the generated sequences with the known sequences of the input design condition and measure their similarity using the recovery ratio. The low recovery ratios listed indicate the generated sequences can be different to the known one. At the same time, the normal mode shapes predicted by our model based on the sequence only (green dash curves) also agree well with the measured ones (blue curves). Corresponding to the various normal mode shapes, the generated proteins show a variety of structures, some of which can be related to the vibrational mode shape.

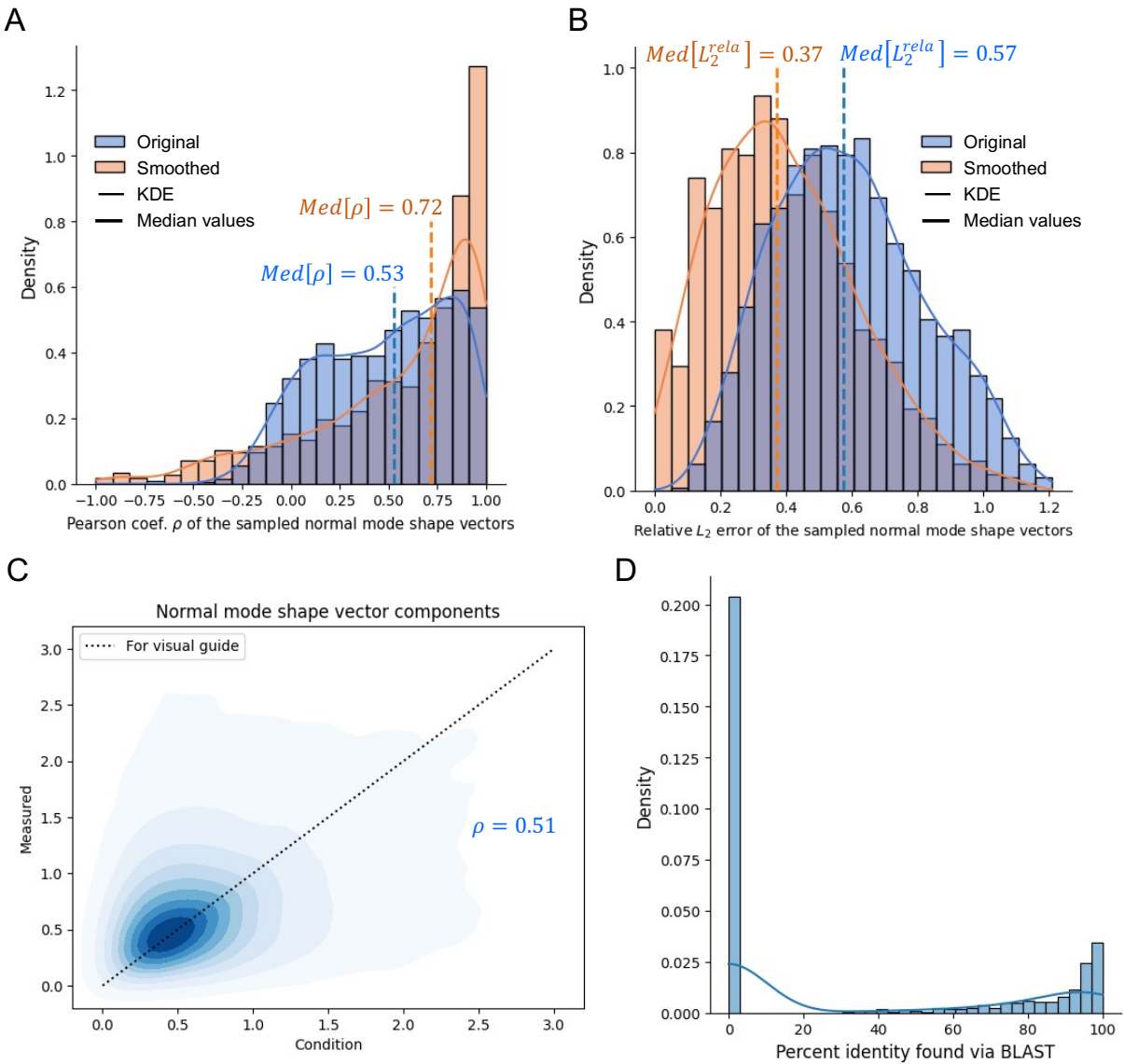

**Fig. 5. Overall quality of generating proteins based on normal mode shapes of existing proteins in the test set.** We test the protein generation model with normal mode shapes from 1,293 proteins in the standalone test set. On the normal mode shape vectors, (A) and (B) show the distribution of Pearson coefficient (A) and relative L2 error (B) in blue for comparing the normal mode shapes of each designed protein with the input conditions while (C) shows the comparison in terms of the components of normal mode shape vectors for all testing cases. By reducing the influence of high-frequency oscillations in the normal mode shape vectors using a low-pass filter and focusing on the low-frequency portions, the Pearson coefficient and relative L2 error of the smooth normal mode vectors are shown in red in (A) and (B). On the novelty of the designed sequences, (F) shows the distribution of the highest percent identity found via BLAST test.

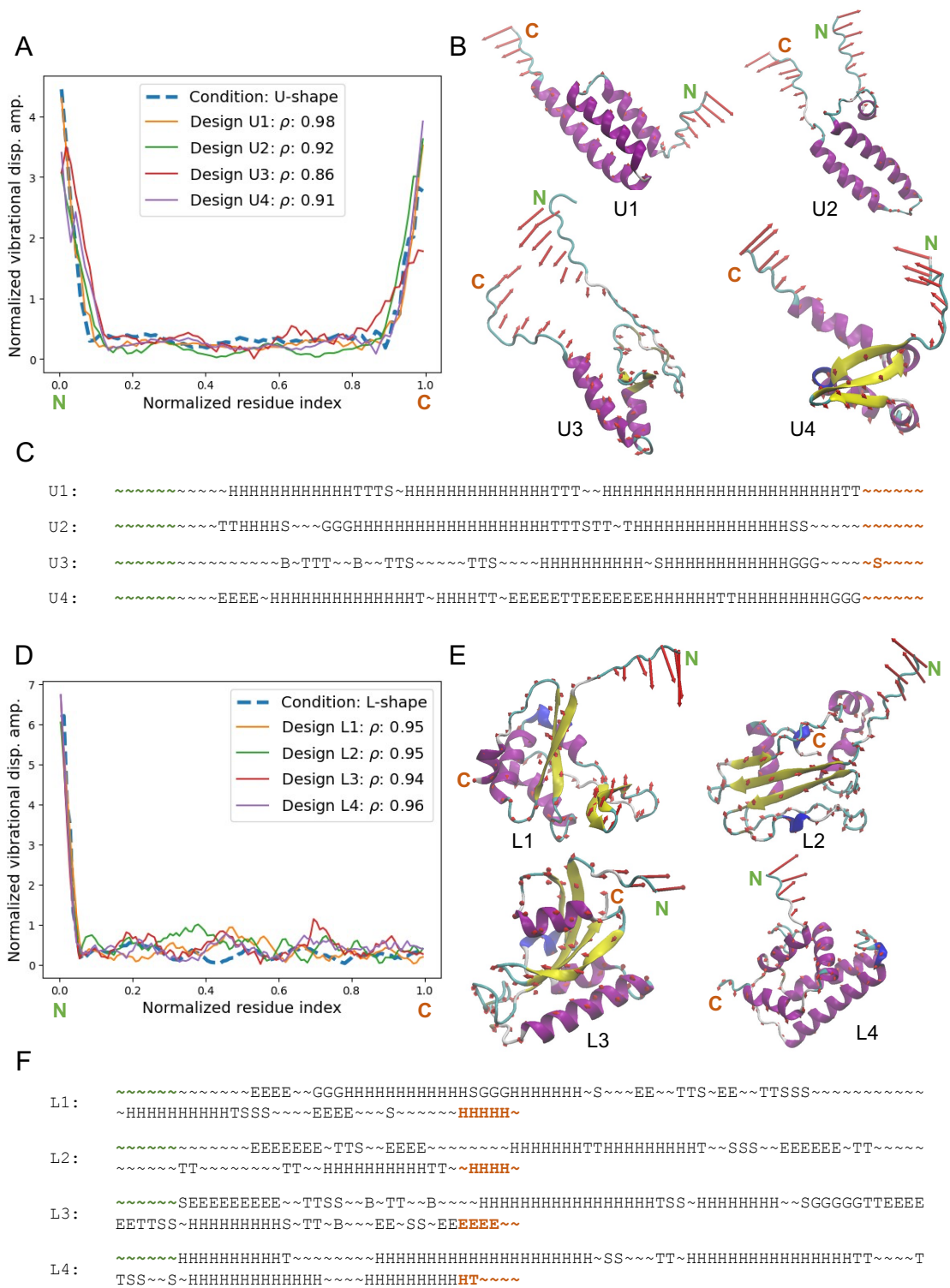

**Fig. 6. Diverse protein sequences generated for the same input normal mode shape vectors.** For a U-shape normal mode shape vector (A-C) and an L-shape condition (D-F), our protein generation model can generate multiple sequences (listed in Table 2) with high design accuracies (A and D). The corresponding 3D protein structures (B and E) and 1D secondary structure sequence (C and F) show similarity and diversity at different locations along the sequences.

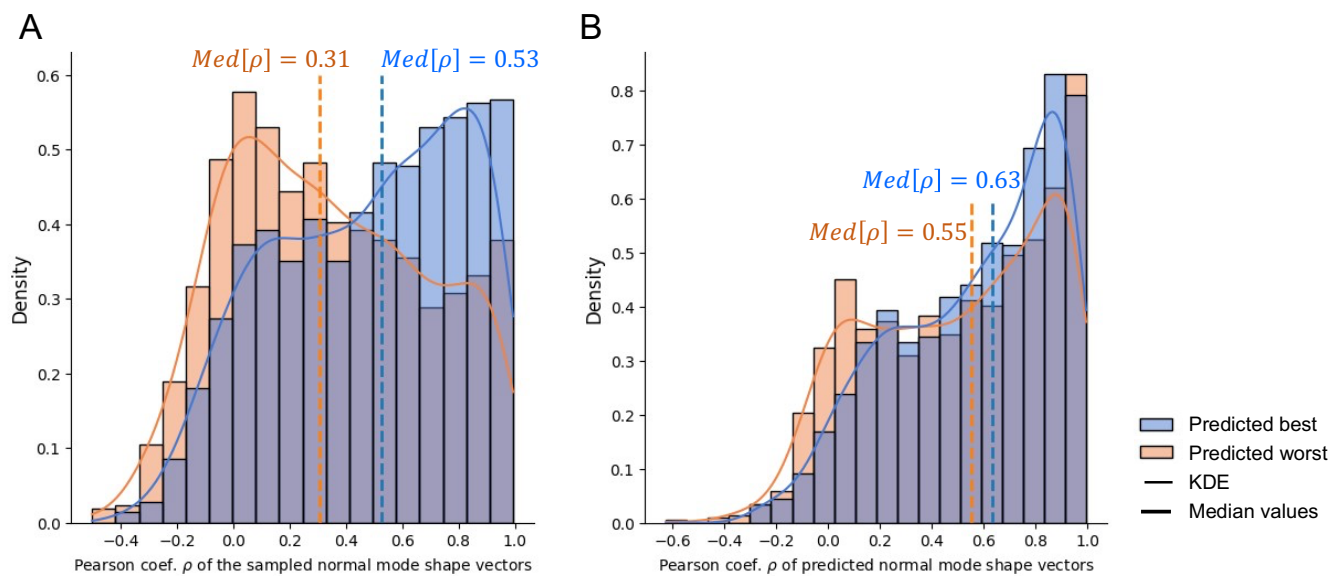

**Fig. 7. Comparing the best and worst design groups identified by the protein predictor.** (A) On the design accuracy in terms of Pearson coefficient validated via normal mode analysis, the predicted best group (in blue) clearly outperforms the predicted worst group (in red) with different distribution shapes and median values. (B) On the predicting accuracy, the protein predictor performs very similarly for the two groups.

**Table 1. Results of the BLAST analysis for the various generated proteins (from Fig. 4) based on normal mode shapes of existing proteins.** Given the normal mode shape vectors of existing proteins as the design condition, the model still yields high probability in predicting sequences that show little similarity to existing proteins as can be seen from the BLAST results (B-E). For other cases, sequences with some similarity to known proteins can be predicted (A and E).

| Case | Sequence | BLAST result: the sequence producing the most significant alignment | |
| --- | --- | --- | --- |
| | | **among PDB proteins** | **beyond PDB proteins** |
| A | MSEDTKKVRCILRRNPIKACKEIKKGNL YKKLPEFKLKEEIPLSIEEKDKNADDAA IQKLLEELTGQETVPEVFIIGGKIGGCT DTVKLYRDGELEPLLREANALL | 59% query cover, 68.25% identical with 3FZ9_A | -- |
| B | MSSGSSGKKLLARYYAVECLVELLKNIV LVSVDLSAQIKRMKEKQGAAFLAVIQLL DQANPGSLEKQGRLPSVLEELQSFARIQ QKDLKAPKFSPDKFSSSSSSGPSSG | -- | No significant similarity found (NSSF) |
| C | GSSGSSGASSAALSIPEKLQTELLAALS EIGISLLNSKSEAKNLLPASLSDKEVQK ISIGVKKRDMKNIKEELEEEGRKSWLAE SLQRQDKKALLVKSNLPPSSNSSSSGPS S | -- | NSSF |
| D | MRRKELETFKSILVIILIFSIAIVVIIY VDDDVKE | -- | NSSF |
| E | MFTTTEVVTVFPGTAVELLVVVSDILPS VASPLKYVTSGLEGEGVVVVENAGGPVV VSCVERITSAGTPGVIEVVVVSGDTQAV ASVGSVSGVAVVELIGYTVALRSRRDVI LVLKFLL | -- | NSSF |
| F | LKCNKLVPLFYKTCPAGKNLCYKMEMVS GGTVIVKRGCIDVCPKSSLLVKYVCCNT DLCNG | -- | 98% query cover, 90.0% identical with P07525.1 |

**Table 2. Results of the BLAST analysis for the multiple protein sequences generated based on the common normal mode shapes (from Fig. 6).**

| Case | Sequence | BLAST result: the sequence producing the most significant alignment | |
|------|----------|---------------------------------------------------------------------|---|
| | | **among PDB proteins** | **beyond PDB proteins** |
| U1 | MSSGSSGGKKKLEELEKELYLSLIPLCP RSIKLACREKIDRRKKEKTRRDKLKSFA KLAIKYERDLNSKIKLSGPSSG | -- | No significant similarity found (NSSF) |
| U2 | MSSGSSGSITAFQLQNDNLDSSCSSLSK VVDLVVQVQSNDLKVLQVRDDDNSTAAL AHTLAEASKQFPVSPSGSGPSS | -- | NSSF |
| U3 | MSSGSSGAKKEVNLGLTCEVCKKDFDEG GELASGPCGEKHKLDCCTELLKKKSKCR REIRAALRRDLDRPSRSGPSSG | -- | 62% query cover, 39.6% identical with KAH7441044.1 |
| U4 | MSSGSSGGVKVRLLSDEENILLVKLLKV AGGRSLLEEIKEKVEGKKKFLIIKLEKI SAIGYEEEKLKKDRKKSGPSSG | -- | NSSF |
| L1 | GSSGSSGGKKTRLVSIEILKKDLSALIQ VVDFVFSEEGKLIIEDILEPRLIKRNKD GITKKKLGEESEALRVPEIKKSGKEQII LEAYKNLNPPSTVSFFTVIKKKKIRIVK EDILK | -- | NSSF |
| L2 | MSHVGSMTLREVNIVLVVIVTPSGSEIE VAGRVELQVNLAKLAAEGSLRVLRILTG SVCAPVGRVLFAVVLPGNRNVGSFRELT PSASLVEIQVQGFDLGLLLKKLFRRGVS LLLLL | -- | NSSF |
| L3 | GSVEEPARVRVSHLLVKHSQSRRPSSSR QEKITRTKEEALELLSGYLKKKKSGEEE FEERASQKSDDSSAKRGGDLGFFSRGQM VKPFEDAAFALKTGEISGPVFTDSGYHI ILRTE | 100% query cover, 82.1% identical with 2RUD_A | -- |
| L4 | GSIMEPARVRVSHLLVKHSKSRRPSSFE KKKITRLKEDLLELFNEIGAEEFKLGSE KDSKLANKAALFFRVINFKFKKGDGVVC GKGSYVAAVLLVTASNVDLEILEEFISS SRKPK | 55% query cover, 59.4% identical with 5GPH_A | -- |

# Agentic End-to-End *De Novo* Protein Design for Tailored Dynamics Using a Language Diffusion Model

Bo Ni[1,2], Markus J. Buehler[1,3,4]*

[1] Laboratory for Atomistic and Molecular Mechanics (LAMM), Massachusetts Institute of Technology, 77 Massachusetts Ave., Cambridge, MA 02139, USA

[2] Department of Materials Science and Engineering, Carnegie Mellon University, 5000 Forbes Avenue, Pittsburgh, PA 15213, USA

[3] Center for Computational Science and Engineering, Schwarzman College of Computing, Massachusetts Institute of Technology, 77 Massachusetts Ave., Cambridge, MA 02139, USA

[4] Lead contact

*Correspondence: mbuehler@MIT.EDU

# SUPPLEMENTARY INFORMATION

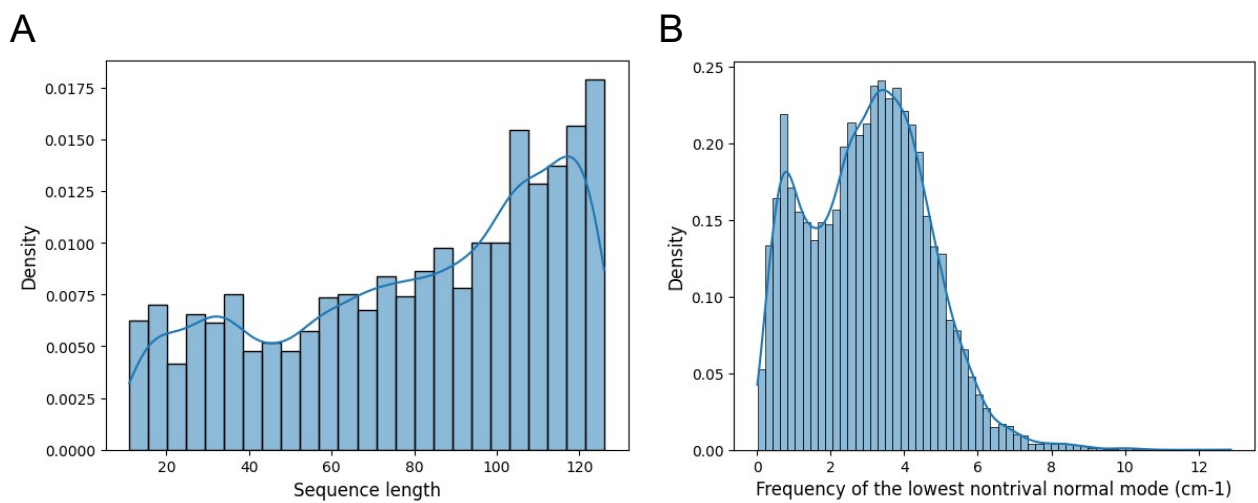

**Fig. S1. Distributions of the lowest non-trivial normal mode dataset of PDB proteins curated.** (A) sequence length and (B) normal mode frequency.

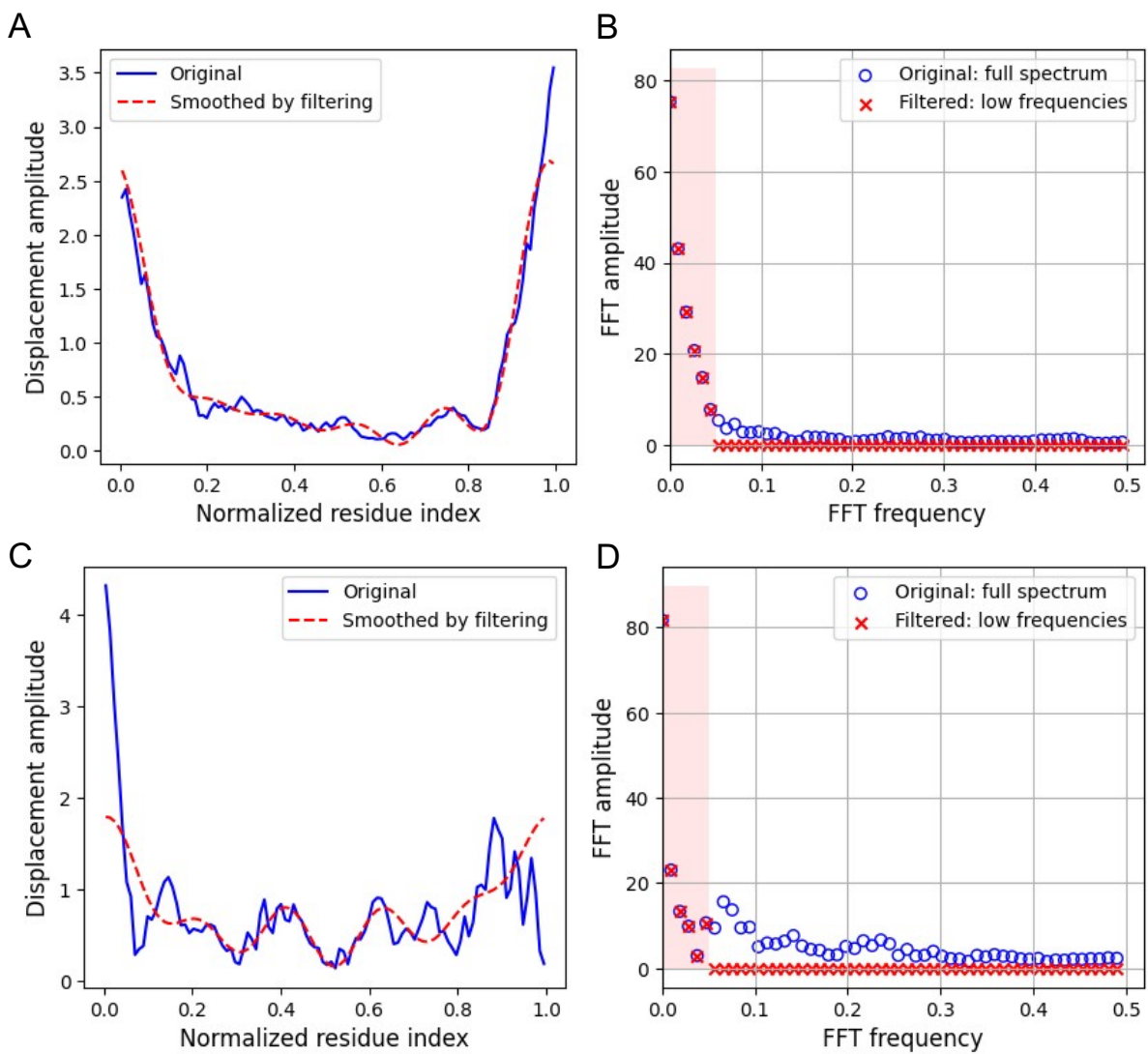

**Fig. S2. Applying a low-pass filter to smooth the normal mode shape vectors.** (A)/(C) showed the original (blue solid lines) and smoothed (red dash lines) normal mode shape vectors in the real space. (B)/(D) shows the corresponding vectors in the frequency domain described as the fast Fourier transformation amplitude over different frequencies. To smooth the vector in real space, only the components with the lowest 10% frequencies (red crosses in B/D) are kept.

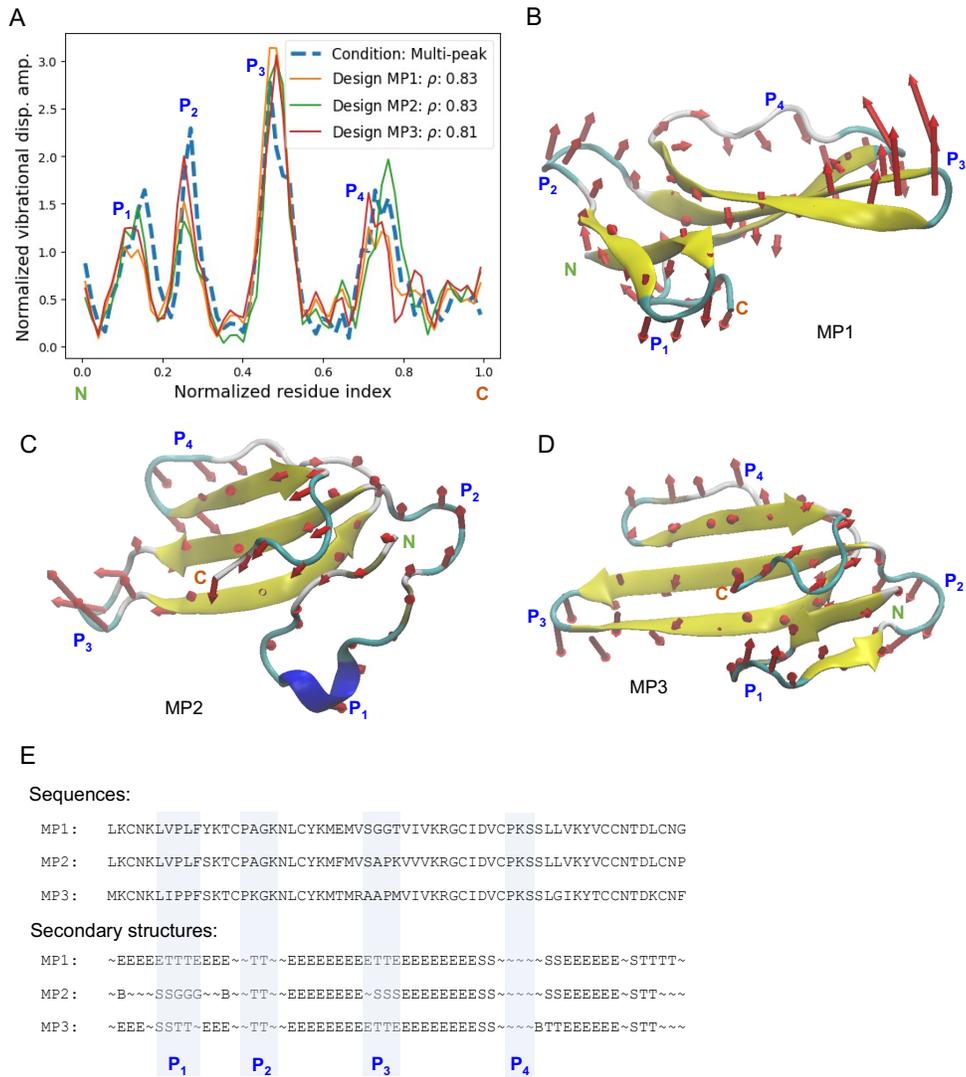

**Fig. S3. Multiple protein sequences generated for the same input normal mode shape vectors.** While these designs all achieve high accuracy (A), their 3D structures as beta-sheets connected with coils and turns (B-D) as well as secondary structure sequences (E) show strong similarities.